Second-order correlation function from asymmetric to symmetric transitions due to spectrally indistinguishable biexciton cascade emission

X. F. Wu,[1] X. M. Dou,[1] K. Ding,[1] P. Y. Zhou,[1] H. Q. Ni,[1] Z. C. Niu,[1] H. J. Zhu,[1] D. S. Jiang,[1] C. L. Zhao,[2] and B. Q. Sun[1*]

[1]*State Key Laboratory for Superlattices and Microstructures, Institute of Semiconductors, Chinese Academy of Sciences, Beijing, 100083, China*
[2]*College of Physics and Electronic Information, Inner Mongolia University for Nationalities, Tongliao 028043, China*

We report the observed photon bunching statistics of biexciton cascade emission at zero time delay in single quantum dots (QDs) by second-order correlation function $g^{(2)}(\tau)$ measurements under continuous wave excitation. It is found that the bunching phenomenon is independent of the biexciton binding energy when it varies from 0.59 meV to nearly zero. The photon bunching takes place when the exciton photon is not spectrally distinguishable from biexciton photon, and either of them can trigger the "start" in a Hanbury-Brown and Twiss setup. However, if the exciton energy is spectrally distinguishable from the biexciton the photon statistics becomes asymmetric and a cross-bunching lineshape is obtained. The theoretical calculations based on a model of three-level rate-equation analysis are consistent with the result of $g^{(2)}(\tau)$ correlation function measurements.

*Electronic address: bqsun@semi.ac.cn

Two-photon cascade emission, simultaneous two-photon emission and related photon statistics in atoms and single quantum dots (QDs) have been studied using photon-correlated measurements [1-6]. The observed cross-correlation function $g^{(2)}(\tau)$ for the spectrally separated exciton (X) and biexciton (XX) photons under continuous wave (CW) excitation is asymmetric at zero time delay, showing an antibunching-to-bunching behavior [1,4]. To enable the observation of simultaneously spontaneous two-photon emission, a high-Q microcavity has been employed to enhance the two-photon transition and at the same time suppress the single-photon cascade decay in atoms and single QDs [7,8], where the cavity mode is tuned to be the half energy of the transition. It is also reported that, even without cavity-enhanced effect, a two-photon emission may exist in single QDs when the biexciton binding energy is less than the full width at half maximum (FWHM) of the QD emission [9]. It is noticed that the analysis was based on the symmetric bunching statistics observed at zero time delay.

However, as is known from atomic physics [3, 10], two-photon transitions are much weaker than the related first-order one-photon process. Hence, in order to observe the second-order spontaneous two-photon process the first-order transition should be either forbidden by selection rules [10,11] or suppressed by a cavity [7,8,10]. Such an effect can also be predicted for optical transition in single QDs in the solid state. Therefore, the question about whether the photon statistics of the biexciton cascade emission will be the dominant process for contributing the bunching peak of $g^{(2)}(\tau=0)$ or not is still needed to investigate theoretically and experimentally under the conditions of different biexciton binding energies and spectral bandpass in the $g^{(2)}(\tau)$ measurements in detail.

In this letter, we demonstrate the symmetric bunching statistics of biexciton cascade emission at zero time delay in single QDs coupled with a low quality factor cavity under CW excitation. The X and XX emission energies, as well as biexciton binding energy，can be continuously tuned using hydrostatic pressure, which helps us to realize the QD emission resonance with the cavity mode and at the same time allows us to investigate the correlation between the bunching phenomenon and  varying biexciton binding energy by using $g^{(2)}(\tau)$ measurements. We found that the photon bunching is due to the spectral indistinguishable exciton and biexciton photons and is independent of the variation of biexciton binding energy. If the X photon is spectrally distinguishable from the XX photon the photon statistics becomes asymmetric and an antibunching-to-bunching lineshape is shown. The observed $g^{(2)}(\tau)$ bunching phenomena can be well simulated based on the rate-equations of a QD three-level scheme.

The studied InAs/GaAs QD sample was grown by molecular-beam epitaxy on a (001) GaAs substrate that consisted of 20/15 (bottom/top) pairs of GaAs/Al$_{0.9}$Ga$_{0.1}$As bilayers as a distributed Brigg reflectors (DBR), and a λ GaAs cavity layer with an InAs QD layer. The QD emissions mainly ranged from 890 to 940 nm and the cavity mode was 870 nm at zero pressure with a low quality factor (Q=λ/Δλ ~ 720 for λ ~ 864 nm and Δλ ~ 1.2 nm, see Fig.1(a)). The QD photoluminescence (PL) wavelength can be continuously adjusted using diamond anvil cell (DAC) at temperature of 20 K to match the cavity mode for enhancing QD emission [12,13] as shown in Fig.1(a), where PL wavelength changes as a function of tuning pressure from 0.51 to 0.60 GPa. The cavity mode was determined using the reflectance spectrum of the sample measured with a halogen lamp as shown in Fig.1(a) by red curves. For optical excitation of PL, a CW diode laser emitting at 640 nm was focused to a spot with a diameter of ~ 4 μm on the sample using a microscope objective with a numerical aperture of 0.35. The PL from the QD was collected by the same objective into a monochromator (SpectraPro-500i, spectral resolution of 0.05 nm) with a focal length of 0.5 m equipped with a silicon charge-coupled device (CCD). In the measurements of $g^{(2)}(\tau)$, the PL signal was sent to a Hanbury-Brown and Twiss (HBT) setup equipped with two monochromators and two silicon avalanche photodiodes (APDs) as shown in Fig. 1(b), whose timing resolution was 300 ps. Three different spectral bandpass configurations which are modulated by changing the slit widths and gratings (1200 or 600 g/mm) of monochromators can be arranged for the cross-correlation measurements of XX (X) and X (XX) by the start (stop) triggering operation of the APDs, respectively. A wide range of spectral bandpass was also employed for PL spectral measurements and simultaneously collecting both X and XX photons during the autocorrelation measurements. In addition, for measuring high-resolution PL (HRPL) spectra, i.e., to exactly obtain FWHM and the energy separation between X and XX PL peaks, a scanning Fabry-Perot interferometer (FPI) with a free spectral range of 15 GHz (62 μeV) and a spectral resolution of 2 μeV was used.

We have earlier reported that the blue shift of the X and XX emission peaks under applied hydrostatic pressure has different rates of 89.85 and 89.06 meV/GPa, respectively [12], and the cavity mode shift rate was 20 meV/GPa [13]. It means that it is able to tune a biexciton binding energy from antibinding to binding state by applying hydrostatic pressure when the X emission energy is less than XX emission energy at zero pressure [14], and the match of the photon energies to the cavity mode can be well modulated. Figure 2 (a) presents the polarization-resolved PL spectra at a pressure of 0.55 GPa, where the QD emission is resonant with the cavity. It clearly shows that the corresponding horizontal (H) polarized X and XX spectra overlap with each other (shown by red line), whereas the vertical (V) polarized X and XX photons have an energy separation of 80 μeV (black line), i.e., the biexciton binding energy $E_B$ is 80 μeV for V-polarization. By using FPI to measure HRPL of H-polarized spectrum, Figure 2 (b) shows that the FWHM of the QD emission lines is approximately 10 μeV, and the separation between X and XX peaks is 15 μeV, i.e., $E_B$~15 μeV for H-polarization. Thus based on the HRPL spectral measurements, H-polarized emission lines are partially overlapped. However, the V-polarized X and XX emissions are essentially distinguishable from each other due to $E_B$ >> FWMH especially if a spectrometer with enough resolution is employed.

Figure 2(c) shows the V- and H-polarized PL spectra measured at three different excitation powers of 21, 40 and 86 μW, which exhibit that the peak wavelengths of both X and XX emissions do not vary with changing excitation power, but the relative PL intensity of X and XX emission lines varies as what is expected. In order to investigate the photon statistics corresponding to the different condition of the biexciton binding energy in H and V configurations, the $g^{(2)}(\tau)$ correlation function measurements are completed at a center wavelength of 865.89 nm with a spectral bandpass of 0.08 nm (the scale is marked in middle part of Fig. 2(c)). For H-polarized PL the center wavelength is located at the peak position, and for V-polarized PL it is at the midpoint between the X and XX emission lines. The power dependence of bunching phenomenon for H- and V-polarized PL by the $g^{(2)}(\tau)$ autocorrelation measurements is presented in Figs. 2(d) and (e), respectively, where the XX or X photon can trigger the "start" in the HBT setup. A reduction of the photon bunching is observed towards the increasing pumping powers. Similar pumping power-dependent effect has been reported for the $g^{(2)}(\tau)$ measurements in single GaN and GaAs QDs [9,15], where the photon bunching of X and XX photons decreases with increasing power, and the phenomenon is shown to concern with the variation of populate probabilities of the exciton, biexciton and charged exciton [3,6,15]. From the result presented in Figs. 2 (d) and (e), it is confirmed that $g^{(2)}(\tau)$ shows a large photon bunching phenomenon and

$g^{(2)}(0)$ value is approximately 3 at the weak pumping power of 21 μW for both H- and V-polarized photons, no matter how large is the energy separation between X and XX emission lines in these two polarization configurations. It implies that the observed bunching behavior is independent of the biexciton binding energy.

To study the correlation between photon bunching phenomenon and spectral bandpass selection during photon-correlated measurements, a QD with better spectral resolved X and XX emission lines as shown in Fig.3 (a) was chosen, where H-polarized X and XX emission lines have a larger separation of 0.36 nm (0.59 meV). During this measurement, a pressure has been applied to shift both X and XX emissions into the cavity mode. The excitation power dependences of unpolarized X and XX emission intensities measured at 64, 120 and 260 μW are shown in Fig. 3(b), respectively, where the spectral resolution for the PL measurements is taken as either 0.08 nm by using a 1200 g/mm grating (red curves) or 1 nm by using a 600 g/mm grating (black curves). Then two typical $g^{(2)}(\tau)$ measurements are completed under two different experimental conditions. One configuration is that the spectral bandpass for both X and XX emission measurements is set to a small value of 0.08 nm (see Fig.3(b)). In this case, cross-correlation measurements can be done when use XX (X) photon to trigger start and X (XX) photon to trigger stop in the HBT setup (see Fig.1 (b)), respectively. Figures 3(c) and (d) present the measured power dependence of $g^{(2)}(\tau)$. The $g^{(2)}(\tau)$ lineshapes show a typical antibunching-to-bunching behavior (in the case of "start" triggered by XX) or a bunching-to-antibunching behavior (in the case of "start" triggered by X) around $g^{(2)}(0)$, respectively [1,4]. In the second configuration of measurement, the spectral resolution is set to 1 nm for the PL measurements, and a broaden PL spectra with a spectral unresolved X and XX emission lines are shown in Fig. 3(b) by black curves. Such a broad bandpass arrangement (see Fig.3(b)) results in an indistinguishable X photon from XX photon in the HBT setup, similar to the case of unresolved V-polarized photons shown in Fig. 2. The $g^{(2)}(\tau)$ is measured as an autocorrelation function. The expected photon bunching of the $g^{(2)}(\tau)$ curve like what observed in Fig. 2 (e) appears again in Fig. 3(f), confirming that the bunching effect is related to spectrally overlapped cascade emissions. It is very interesting to find that a similar bunching result, as shown in Fig.3 (e), can be obtained by only adding two cross-correlation $g^{(2)}(\tau)$ data curves of the Figs. 3(c) and (d). Therefore, such a bunching signature provides strong evidence to the assignment that the autocorrelation function obtained with a broad bandpass of 1 nm really comes from the two real coincident events of the XX cascade process. It is noted that, in the case of strong coupling with high Q nanocavity as reported by Ota et al [7] two-photon emission contribution is about 10% of the whole emission at the two-photon resonance. As a contrast, here in this work, the QDs is located in a low Q cavity, and the cavity effect is mainly to increase both X and XX emission intensities. Actually, no any PL peak corresponding to the two-photon emission, for example, at the midpoint between X and XX peaks, is observed (see Fig.3(a)). Therefore, the contribution from two-photon emission to the obtained $g^{(2)}(\tau)$ result can be neglected.

To numerically calculate photon bunching effect of the QD emissions and analyze the resulted $g^{(2)}(\tau)$ lineshape, we performed a rate-equation analysis using three-level scheme as depicted in Fig. 4 (a), where the QD states are consisted of the biexciton level 2, exciton level 1 and ground level 0. The QD is excited into level 2 at a rate R, where it decays to the ground level through intermediate level 1 by radiative rate $\gamma_2$ and then by a rate of $\gamma_1$ ($\gamma_2 \approx 2\gamma_1$ [15,16]), respectively. The population dynamics is then [1,15,17]

$$d\rho_{22}/dt = R - \gamma_2 \rho_{22}$$
$$d\rho_{11}/dt = \gamma_2 \rho_{22} - \gamma_1 \rho_{11} \qquad (1)$$
$$d\rho_{00}/dt = -R + \gamma_1 \rho_{11}$$

where $\rho_{ii}$ denotes the diagonal density-matrix element of population probability of the level $i$, and $\rho_{00} + \rho_{11} + \rho_{22} = 1$ for weak pumping intensity. $g^{(2)}(\tau)$ correlation functions are obtained in terms of single-time expectation values,

$$g^{(2)}_{22}(\tau) = g^{(2)}_{12}(\tau) = 1 - \exp(-\gamma_2 \tau)$$
$$g^{(2)}_{11}(\tau) = \{\gamma_1[1-\exp(-\gamma_2 \tau)] - \gamma_2[1-\exp(-\gamma_1 \tau)]\}/(\gamma_1 - \gamma_2) \qquad (2)$$
$$g^{(2)}_{21}(\tau) = g^{(2)}_{11}(\tau) + (\gamma_1/R)\exp(-\gamma_1 \tau)$$

where $g^{(2)}_{11}(\tau)$ and $g^{(2)}_{22}(\tau)$ are autocorrelation functions of X and XX one photon process, respectively, $g^{(2)}_{21}(\tau)$ corresponds to the process in which the XX photon emits first and then X photon emission is followed, and $g_{12}^{(2)}(\tau)$ corresponds to the process with a reverse order in time. Therefore, $g_{21}^{(2)}(\tau)$ is responsible for cross-bunching peak as shown in Figs.3 (c) and (d), whereas $g_{12}^{(2)}(\tau)$ corresponds to the antibunching in the cross-correlation measurements. As demonstrated in the reported photon cross-correlation measurements, FWHM of the antibunching is much broader than that of corresponding bunching peak [1,4,18]. This is attributed to the spectral diffusion [18,19], and $g_{12}^{(2)}(\tau)$ function should be revised to include the spectral diffusion term and can be rewritten as $g_{12sd}^{(2)}(\tau) = [1-\exp(-2\gamma_1\tau)][1-\exp(-\gamma_d\tau)]$, where $\tau_d(=\gamma_d^{-1})$ is a diffusion time [18]. Using $g_{21}^{(2)}(\tau)$ and $g_{12sd}^{(2)}(-\tau)$ functions to fit the experimental data in Fig. 3(c) at a pumping power of 64 μW for both positive and negative $\tau$, respectively. The obtained diffusion time $\tau_d$, exciton lifetime $\tau_1(=\gamma_1^{-1})$, and the ratio of $\gamma_1/R$ are 2.50 ns, 0.65 ns, and 5.19, respectively. According to the rate-equation of (1), the steady-state populations of levels 2 and 1 are set equal to $R/\gamma_2(=\rho_{22})$ and $R/\gamma_1(=\rho_{11})$ respectively, which corresponds to $\rho_{22}(0) = 0.1$ and $\rho_{11}(0) = 0.2$. For the ground state, the derived $\rho_{00}(0) = 0.7$, and it remains very close to unity at weak excitation condition. Figures 4 (b) and (c) depicts the calculated cross-correlation function $g^{(2)}(\tau)$ in two different cases, depending on whether XX photon or X photon is used to trigger start in the HBT setup for the cross-correlation $g^{(2)}(\tau)$ measurements. In the calculation, the used parameters are $\gamma_1$=1.54 ns$^{-1}$, $\gamma_d$=0.4 ns$^{-1}$, and $\gamma_1/R$=5.19, and in case (b): when $\tau > 0$, $g_{21}^{(2)}(\tau)$ and $\tau < 0$, $g_{12sd}^{(2)}(-\tau)$; while in case (c): when $\tau > 0$, $g_{12sd}^{(2)}(\tau)$ and $\tau < 0$, $g_{21}^{(2)}(-\tau)$. In addition, for making a comparison with experiment, the corresponding data of case (b) are added to case (c), the result generates a bunching peak at zero time delay as shown in Fig. 4(d). Such a bunching peak is quite similar to what experimentally observed in Figs. 3(e) and (f). The agreement between theory and experiment confirms that the two-photon cascade emission, instead of simultaneous two-photon emission, is responsible for symmetric bunching statistics observed when an enough large spectral bandpass is used during the $g^{(2)}(\tau)$ correlation measurements.

In summary, we have demonstrated that photon bunching signature in the $g^{(2)}(\tau)$ correlation function is due to biexciton single-photon emission through biexciton-exciton cascade. Such bunching phenomenon is independent of the variation of biexciton binding energy, which is evidenced by an experimental tuning of the biexciton binding energy from 0.59 meV to nearly zero. The different photon statistics, from an asymmetric antibunching and bunching phenomenon in the cross-correlation to a symmetric bunching effect in the autocorrelation functions, are determined by changing spectral bandpass in the HBT setup. The observed $g^{(2)}(\tau)$ correlation spectra can be well produced based on the rate-equations of a QD three-level scheme, where the simultaneously spontaneous two-photon emission is actually not generated in the weakly coupled QD-cavity system.

This work was supported by the National Key Basic Research Program of China (Grant Nos. 2013CB922304 and 2013CB933304), the National Natural Science Foundation of China (Grant Nos. 11474275 and 11374295 ).

Figure captions:

FIG.1. (color online) (a) Pressure tuned QD PL emission resonance with the cavity mode. The applied pressure is changed from 0.51 GPa to 0.60 GPa. (b) Hanbury-Brown and Twiss (HBT) setup for $g^{(2)}(\tau)$ measurements, where the adjustment of slit width (i.e. spectral bandpass) of the monochromators located in front of APDs is schematically shown. Three different spectral bandpass configurations are obtained for the XX, X or X+XX detection.

FIG.2. (color online) (a) Polarization-resolved PL spectra for X, XX and $X^+$ emissions. (b) High-resolution H-polarized PL spectra for measuring FWHM and XX binding energy. Blue lines are fitting to data using two Gaussian functions. (c) Polarization-resolved PL spectra at excitation powers of 21, 40, and 86 μW, respectively. Spectral bandpass of 0.08 nm is used for $g^{(2)}(\tau)$ correlation function measurements in either H- (d) or V- (e) polarization.

FIG.3. (color online) (a) Polarization-resolved PL spectra for X, XX and $X^+$ emissions with a peak separation of 0.36 nm between X and XX emission lines for H-polarization. (b) Unpolarized PL spectra measured at excitation powers of 64, 120 and 260 μW with spectral resolutions of 0.08 nm obtained using 1200 g/mm grating (red curves) and 1 nm with 600 g/mm grating (black curves), respectively. The antibunching-to-bunching or bunching-to-antibunching cross-correlation functions measured with a spectral bandpass of 0.08 nm with XX or X to trigger start are shown in (c) and (d), respectively, while autocorrelation function measured with 1 nm bandpass as shown in (f). Figure 3(e) shows a result obtained by adding the corresponding $g^{(2)}(\tau)$ data of (c) and (d). Red lines in (c) are fitting to data using $g^{(2)}(\tau)$ functions, and $\tau_1$ and $\tau_d$ correspond to the exciton lifetime and spectral diffusion time, respectively.

FIG.4 (color online) (a) Energy level scheme of QD states used for the rate equation model. Calculated $g^{(2)}(\tau)$ correlation functions for cases (b): $\tau > 0, g^{(2)}_{21}(\tau); \tau < 0, g^{(2)}_{12sd}(-\tau)$, and (c): $\tau > 0, g^{(2)}_{12sd}(\tau); \tau < 0, g^{(2)}_{21}(-\tau)$. (d) An adding of the corresponding data in (b) and (c) generates a bunching peak at zero time delay.

FIG.1

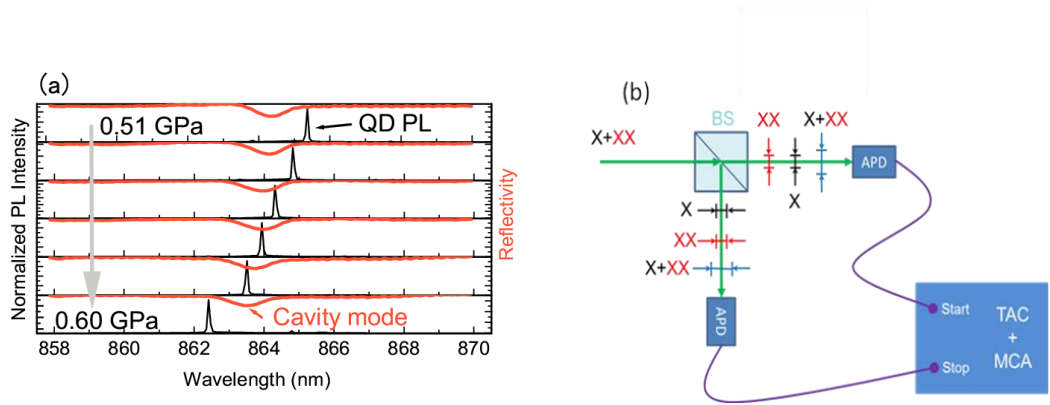

FIG. 2

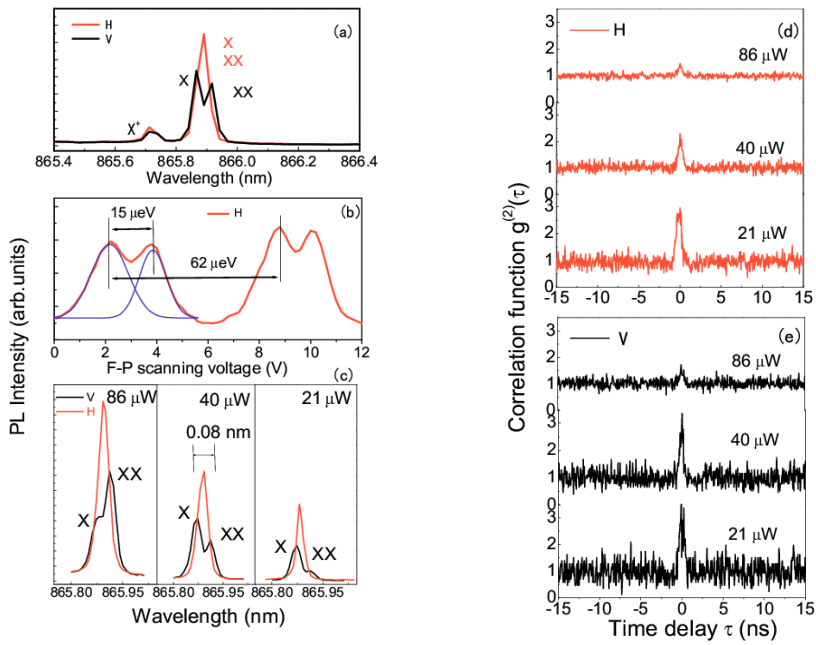

FIG.3

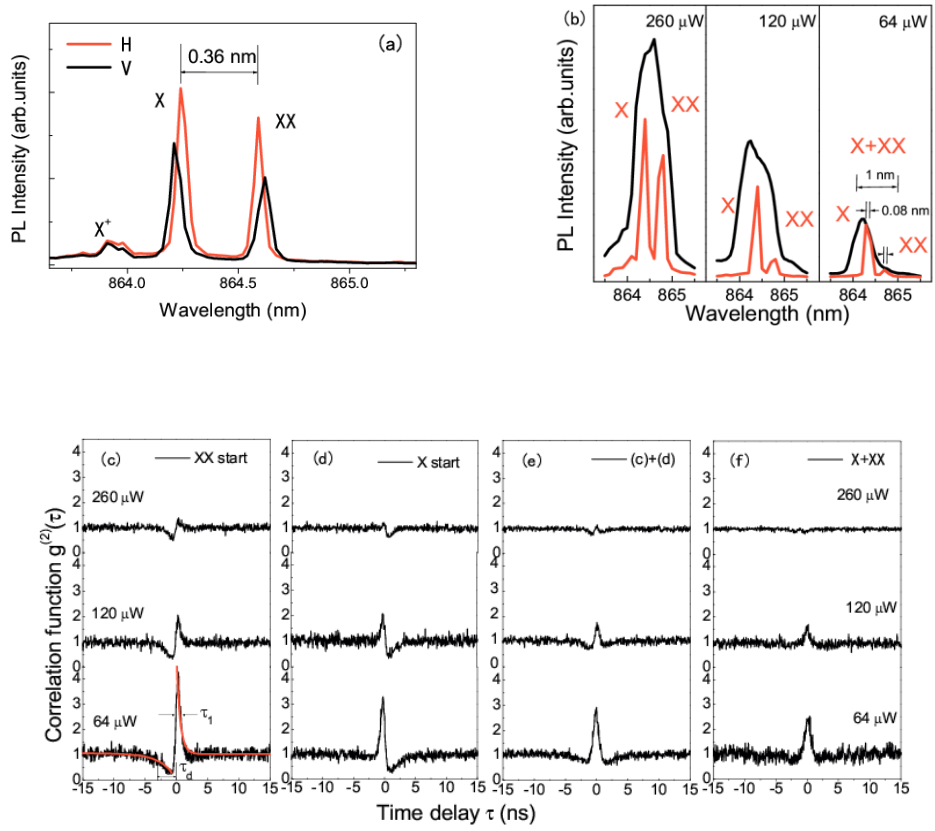

FIG.4

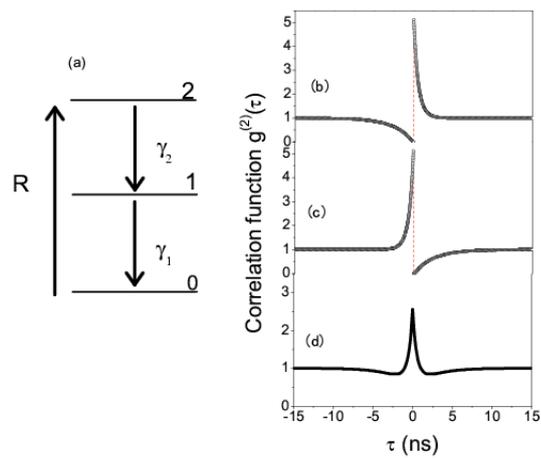